\documentclass[reprint,showpacs,preprintnumbers,amsmath,amssymb,longbibliography,aps,pra]{revtex4-1}
\usepackage[utf8]{inputenc}
\usepackage{mathrsfs}
\usepackage{amsmath}
\usepackage{bm}
\usepackage{amsfonts}
\usepackage{amssymb}
\usepackage{amsthm}
\usepackage{color}
\usepackage{graphicx}
\graphicspath{{./images/}}
\usepackage{geometry}
\usepackage{hyperref}
\usepackage{ulem}
\usepackage{epstopdf}
\usepackage{graphicx}
\hypersetup{
     unicode=false,
     pdftoolbar=true,
     pdfmenubar=true,
     pdffitwindow=false,     
     pdfstartview={FitH},    
     pdftitle={My title},    
     pdfauthor={Author},     
     pdfsubject={Subject},   
     pdfcreator={Creator},   
     pdfproducer={Producer}, 
     pdfkeywords={keyword1} {key2} {key3}, 
     pdfnewwindow=true,      
     colorlinks=true,       
     linkcolor=blue,          
     citecolor=blue,        
     filecolor=magenta,      
     urlcolor=blue           
}
\usepackage[toc,page]{appendix}

\begin{document}
\title{Goos-H\"{a}nchen effect of spin waves at heterochiral interfaces}
\author{Zhenyu Wang}
\author{Yunshan Cao}
\author{Peng Yan}
\email[Corresponding author: ]{yan@uestc.edu.cn}
\affiliation{School of Electronic Science and Engineering and State Key Laboratory of Electronic Thin Films and Integrated Devices, University of Electronic Science and Technology of China, Chengdu 610054, China}

\begin{abstract}
We theoretically investigate the Goos-H\"{a}nchen (GH) effect of spin-wave beams reflected from the interface between two ferromagnetic films with different Dzyaloshinskii-Moriya interactions (DMIs). The formula of the GH shift as functions of the incident angle and material parameters is derived analytically. We show that the GH effect occurs only when spin waves are totally reflected at the interface and vanishes otherwise. We further explore the GH shift of spin waves by narrow DMI strips of different widths. It is found that the induced shift is independent of the strip width down to $10$ nm, offering a novel approach to measure the DMI strength of ultra-narrow magnetic strips which is out the scope of current technology. Full micromagnetic simulations compare well with our theoretical findings. Strong distortion of edge magnetizations for narrower strips however generates a width dependence of the GH shift. The results presented in this work are helpful for understanding the GH effect in chiral magnets and for quantifying the DMI parameter in magnetic strips of sub-$50$ nm scales.
\end{abstract}

\maketitle
\section{Introduction}\label{sec1}
The Goos-H\"{a}nchen (GH) effect originally is a fundamental optical phenomenon in which a light beam reflecting from an interface is laterally shifted \cite{Goos1947}, but it is not limited to optics and can be regarded as a general property of waves, such as acoustics \cite{Declercq2008}, electronics \cite{Chen201301}, and neutron waves \cite{Haan2010}. For spin waves (or magnons), it has been shown that the GH shift can exist when reflecting from the interface between two ferromagnetic films or the edge of a single ferromagnet \cite{Dadoenkova2012,Gruszecki2014,Gruszecki2015}, the properties of which are crucial for the lateral shift of the spin-wave beam. The magnonic GH shift can be a useful tool in characterizing interfaces and edges of magnetic films.

The Dzyaloshinskii-Moriya interaction (DMI) \cite{Dzyaloshinsky1958,Moriya1960}, present in magnetic materials with broken inversion symmetry, has a chiral character and causes the nonreciprocal propagation of spin waves \cite{Zakeri2010,Landeros2013,Moon2013,Garcia2014}, which provides additional functionalities in magnonic devices \cite{Lan2015,Yu2016,Xing2016,Kim2016,Bracher2017}. Recent works found that magnonic crystals with a periodic DMI can efficiently modulate spin-wave propagations and give rise to a plethora of unique effects, such as the spin-wave amplification \cite{Lee2017}, the emergence of indirect gaps, the formation of flat bands, and an unconventional evolution of the standing spin waves around the gaps \cite{Gallardo2019}. Recently, it has been shown that total reflections and negative refractions can occur at the DMI interface and the spin-wave refraction is not symmetric for positive and negative incident angles \cite{Wang2018,Mulkers2018}. However, an in-depth understanding of spin-wave propagations at heterochiral interfaces is still lacking, particularly the magnonic GH effect at the DMI interface.

In this work we investigate theoretically the GH shift at heterochiral interfaces with inhomogeneous DMIs, which can be realized in experiments via engineering the substrate and/or the capping layer of thin ferromagnetic films \cite{Chen201310,Torrejon2014,Wells2017,Tacchi2017}. Here we focus on the high-frequency spin waves, in which the influence of the dipolar interaction can be neglected. We shown that the GH shift occurs only in the case of total reflection and disappears otherwise. Theoretical results are verified by full micromagnetic simulations. We further explore the GH shift at the DMI interface of a narrow magnetic strip. We demonstrate that the GH shift there is the same as that in heterochiral films, i.e., independent on the strip width down to $10$ nm. Our findings promote the heterochiral interface as an advanced platform for spin wave manipulations and enable us to measure the DMI for ultra-narrow (sub-$50$ nm) magnetic strips via the magnonic GH shift, which fills in the gap of current technology.

The paper is organized as follows. In Sec. \ref{sec2}, we present the theoretical model describing the spin-wave propagation. An effective Schr\"{o}dinger equation for spin waves  is established to obtain spin-wave boundary conditions at the DMI interface. The analytical formula of the reflectance and the GH shift is derived. Section \ref{sec3} gives the results of micromagnetic simulations to verify theoretical predictions. Discussion and conclusion are drawn in Sec. \ref{sec4} and \ref{sec5}, respectively.

\section{Analytical Model}\label{sec2}
\begin{figure}
  \centering
  \includegraphics[width=0.5\textwidth]{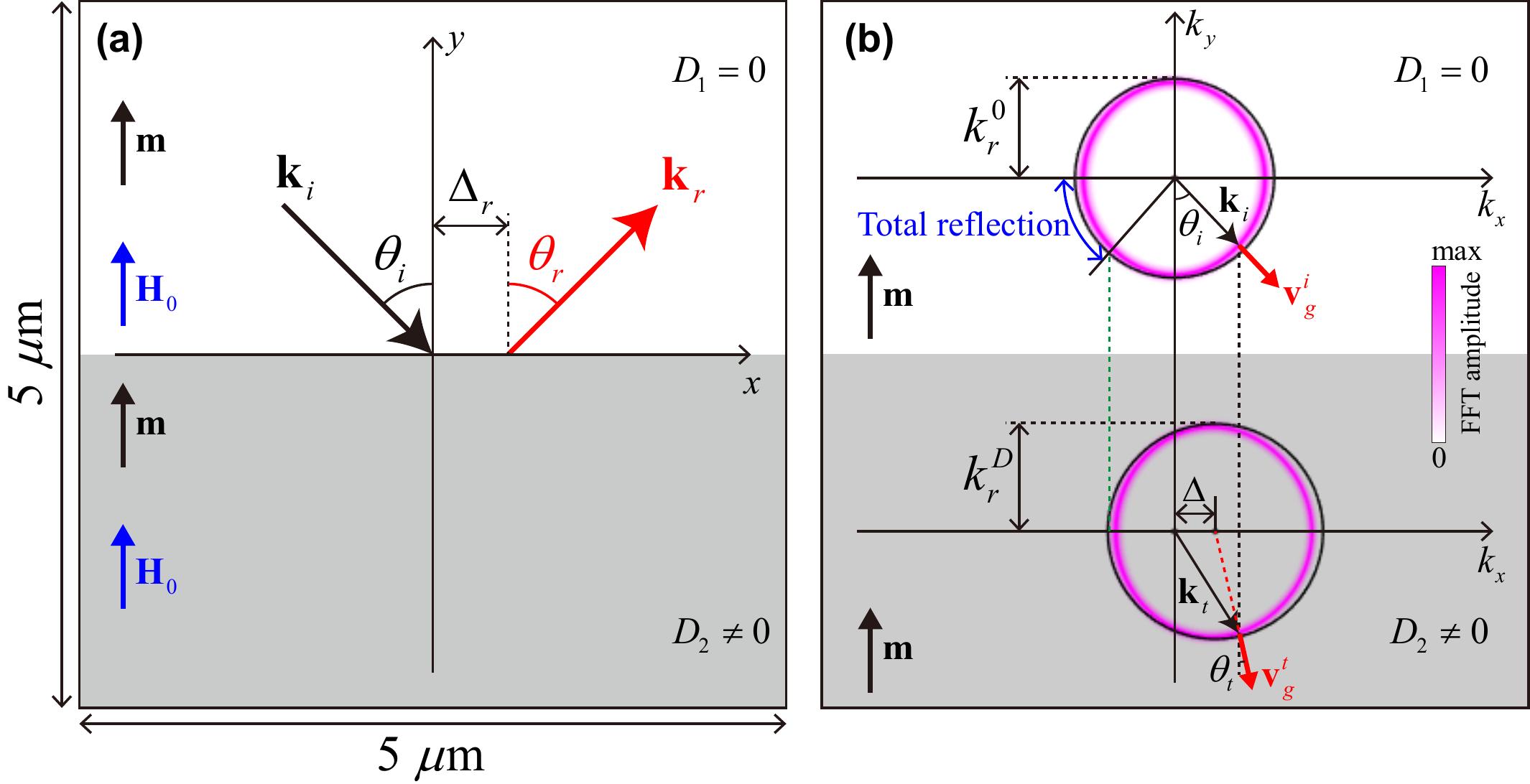}\\
  \caption{(a) Schematic plot of the GH shift at heterochiral interfaces. The external magnetic field and the static magnetization are both oriented along $y$ axis. $\mathbf{k}_{i}$ and $\mathbf{k}_{r}$ denote the wave vectors of the incident and reflected spin-wave beams, $\theta_{i}$ and $\theta_{r}$ are the incident and reflected angles, respectively. $\Delta_{r}$ is the GH shift of the spin-wave beam reflected at the DMI interface. (b) Fast Fourier transformation (FFT) of the simulated spin waves excited by a point source with $\omega/2\pi=100$ GHz in no-DMI and DMI regions, respectively. The upper (lower) circle represents the the isofrequency curve for spin waves  propagating in no-DMI(DMI) region, based on Eq. (\ref{eq_dispersion}). The blue two-way arrow indicates the spin-wave modes which are totally reflected due to the lack of propagation modes with the same $k_{x}$ across the DMI interface.}\label{fig1}
\end{figure}

We first consider the GH shift at the interface of heterochiral magnetic films, i.e., $D_{1}=0$ for $y>0$ and $D_{2}\neq0$ for $y<0$, as shown in Fig. \ref{fig1}(a). To capture the chiral effect solely, we assume the other magnetic parameters to be the same in the heterochiral system. The dynamics of spin-wave propagation in magnetic films is governed by the Landau-Lifshitz-Gilbert (LLG) equation
\begin{equation}\label{eq_llg}
    \frac{\partial\mathbf{m}}{\partial{t}}=-\gamma\mu_{0}\mathbf{m}\times\mathbf{H}_{\mathrm{eff}}+\alpha\mathbf{m}\times\frac{\partial\mathbf{m}}{\partial{t}},
\end{equation}
where $\mathbf{m}=\mathbf{M}/M_{s}$ is the unit magnetization vector with the saturation magnetization $M_{s}$, $\gamma$ is the gyromagnetic ratio, $\mu_{0}$ is the vacuum permeability, and $\alpha$ is the Gilbert damping constant. The effective field $\mathbf{H}_{\mathrm{eff}}$ comprises the exchange field, the DM field, the external field, and the demagnetization field. The DMI considered here has the interfacial form \cite{Bogdanov2001}:
\begin{equation}\label{eq_dmi}
  \mathbf{H}_{\mathrm{DM}} = \frac{2D}{\mu_{0}M_{s}}[\nabla{m}_{z}-(\nabla\cdot\mathbf{m})\hat{z}],
\end{equation}
where $D$ is the DMI constant. An in-plane static external field $\mathbf{H}_{\mathrm{ext}}=H_{0}\hat{y}$ is applied to saturate the magnetization in the film plane ($\mathbf{m}_{0}=+\hat{y}$), see Fig. \ref{fig1}(a).
We assume a small fluctuation of $\mathbf{m}$ around $\mathbf{m}_{0}$, and express the magnetization as $\mathbf{m}=m_{0}\hat{y}+m_{x}\hat{x}+m_{z}\hat{z}$ with $m_{0}\approx1$ and $m_{x,z}\ll1$. The exchange spin wave with high frequency is considered and the dipolar interaction is ignored (wave lengths below 100 nm \cite{Lenk2011}), so that the spin-wave dispersion relation reads \cite{Wang2018}
\begin{equation}\label{eq_dispersion}
  \omega=A^{\ast}\mathbf{k}^{2}+\omega_{H}-D^{\ast}k_{x},
\end{equation}
where $A^{\ast}=2\gamma{A}/M_{s}$ with the exchange constant $A$, $\omega_{H}=\gamma\mu_{0}H_{0}$, $D^{\ast}=2\gamma{D}/M_{s}$, and $\mathbf{k}=(k_{x},k_{y})$ is the wave vector of the spin wave. The spin-wave propagation characteristic can be analyzed by the isofrequency curve, as illustrated in Fig. \ref{fig1}(b). In the region 1 without the DMI ($D=0$), the spin-wave isofrequency curve at a given frequency $\omega$ is a circle centered at the origin with the radius $k_{r}^{0}=\sqrt{(\omega-\omega_{H})/A^{\ast}}$. In the presence of the DMI ($D\neq0$), the isofrequency circle is shifted by $\Delta=D^{\ast}/2A^{\ast}$ along the $+k_{x}$ axis and its radius increases to $k_{r}^{D}=\sqrt{(k_{r}^{0})^{2}+\Delta^{2}}$. Micromagnetic simulations agree well with the analytical formula Eq. (\ref{eq_dispersion}) [see Fig. \ref{fig1}(b)], which justifies the validity of the spin-wave dispersion relation.

The scattering behaviors of spin waves at the heterochiral interface can be described by the generalized Snell's law \cite{Wang2018,Mulkers2018}:
\begin{equation}\label{eq_snell}
  k_{r}^{0}\sin\theta_{i}=k_{r}^{D}\sin\theta_{t}+\Delta,
\end{equation}
where $\theta_{i}$ and $\theta_{t}$ are the incident and refracted angles of spin-wave beams, respectively, with respect to the interface normal ($y$ axis), see Fig. \ref{fig1}(b).
According to Eq. (\ref{eq_snell}), we obtain the critical angle for total reflection,
\begin{equation}\label{eq_thtic}
  \theta_{ic}=\arcsin\Big[\frac{\Delta-k_{r}^{D}}{k_{r}^{0}}\Big],
\end{equation}
corresponding to $\theta_t=-90^{\circ}$.
The dependence of the critical angle on the DMI constant is plotted in Fig. \ref{fig2}(b) [dashed line]. One can see that $\theta_{ic}$ increases with $D$. Total reflection occurs only when the incident angle satisfies $\theta_{i}<\theta_{ic}$.

The evaluation of the reflection and transmission coefficients requires boundary conditions for spin waves at the heterochiral interface, which are derived below.
By defining a spin-wave function $\Phi=m_{x}+{i}m_{z}$ and neglecting the damping, we linearize the LLG equation (\ref{eq_llg}) and recast it into an effective Schr\"{o}dinger equation \cite{Lee2017}:
\begin{equation}\label{eq_schrodinger}
  i\hbar\frac{\partial\Phi}{\partial{t}}=\Big[\frac{\hat{p}^{2}}{2m^{\ast}}+\gamma\hbar\mu_{0}H_{0}-D^{\ast}(y)\hat{p}_{x}\Big]\Phi,
\end{equation}
where $\hat{p}=-i\hbar\nabla$ is the momentum operator and $m^{\ast}=\hbar/2A^{\ast}$ is the effective mass of spin waves. From the continuity of the wave function and integrating Eq. (\ref{eq_schrodinger}) over the vicinity of the interface, we obtain the boundary conditions for the magnon wave function $\Phi$ at the DMI interface between two regions as
\begin{equation}\label{eq_bc}
\begin{split}
  &\Phi_{1}(y=0) = \Phi_{2}(y=0), \\
  &\frac{d\Phi_{1}}{dy}\Big|_{y=0}-\frac{d\Phi_{2}}{dy}\Big|_{y=0} = 0.
\end{split}
\end{equation}

The incident and reflected waves in region 1 and the refracted wave in region 2 are assumed as plane waves:
\begin{equation}\label{eq_wavefuction}
  \begin{split}
    \Phi_{1} &=e^{i(\mathbf{k}_{i}\cdot\mathbf{r}-\omega{t})}+{R}e^{i(\mathbf{k}_{r}\cdot\mathbf{r}-\omega{t})}, \\
    \Phi_{2} &={T}e^{i(\mathbf{k}_{t}\cdot\mathbf{r}-\omega{t})},
  \end{split}
\end{equation}
where $R$ and $T$ are the reflection and transmission coefficients, respectively. Due to the translational symmetry along the interface, the wave vector component tangential to the interface is conserved $k_{ix}=k_{rx}=k_{tx}$. Moreover, in region 1 where $D=0$, the dispersion relation is isotropic which implies $k_{iy}=-k_{ry}$. A spin-wave beam incident at an angle $\theta_{i}$ has the wave vector $\mathbf{k}_{i}=k_{ix}\hat{x}+k_{iy}\hat{y}$ with $k_{ix}=k_{r}^{0}\sin\theta_{i}$ and $k_{iy}=-k_{r}^{0}\cos\theta_{i}$. We therefore obtain the wave vector of the refracted spin wave $\mathbf{k}_{t}=k_{tx}\hat{x}+k_{ty}\hat{y}$ with $k_{tx}=k_{r}^{0}\sin\theta_{i}$ and $k_{ty}=-\sqrt{(k_{r}^{D})^{2}-(k_{r}^{0}\sin\theta_{i}-\Delta)^{2}}$.

Substituting Eq. (\ref{eq_wavefuction}) into the boundary conditions Eq. (\ref{eq_bc}) gives the reflection and transmission coefficients:
\begin{equation}\label{eq_RT}
  \begin{split}
    R &= \frac{k_{iy}-k_{ty}}{k_{iy}+k_{ty}}, \\
    T &= \frac{2k_{iy}}{k_{iy}+k_{ty}}.
  \end{split}
\end{equation}
To calculate the GH shift, we use the stationary phase method \cite{Artmann1948}, which is based on the phase difference between the reflected and incident beams.
If the incident spin wave is a wave packet of a Gaussian shape with the $x$ component of the wave vector variation $\Delta{k_{x}}\ll{k_{x}}$, the GH shift of the reflected beam is given by Artmann's formula \cite{Artmann1948,Gruszecki2015,Gruszecki2017}:
\begin{equation}\label{eq_GHr}
  \Delta_{r}=-\frac{\partial\varphi_{r}}{\partial{k_{x}}}\mathrm{sgn}(\theta_{i}),
\end{equation}
where $\varphi_{r}=\arctan[\mathrm{Im}(R)/\mathrm{Re}(R)]$ is the phase difference between the reflected and incident waves, with Im($R$) and Re($R$) being the imaginary and real parts of the reflection coefficient, respectively,  calculated from Eq. (\ref{eq_RT}).
The sign function $\mathrm{sgn}(\theta_{i})$ is defined as follows: when a spin-wave beam is incident from left (right) side, $\theta_{i}$ is positive (negative) and $\mathrm{sgn}(\theta_{i})=1 (-1)$.

In the case of total reflection ($\theta_{i}<\theta_{ic}$), $k_{ty}$ is imaginary and $R$ is complex. The GH shift for this case is given as
\begin{equation}\label{eq_GHrTR}
  \Delta_{r}=\frac{2(1+\sin^{2}\theta_{i})}{\sin2\theta_{i}\sqrt{(k_{r}^{0}\sin\theta_{i}-\Delta)^{2}-(k_{r}^{D})^{2}}}\mathrm{sgn}(\theta_{i}).
\end{equation}
For the other incident angles ($\theta_{i}>\theta_{ic}$), both $k_{ty}$ and $R$ are real numbers, which leads to $\psi_{r}=0$ and $\Delta_{r}=0$.
The nonreciprocal GH shift [$\Delta_{r}(\theta_{i})\neq\Delta_{r}(-\theta_{i})$] occurs because the isofrequency is shifted in $k$ space for the DMI region [the lower part in Fig. \ref{fig1}(b)], which is embodied in the generalized Snell's law [Eq. (\ref{eq_snell})] by the additional term $\Delta$. This shift can also cause the asymmetric refraction of spin waves at the heterochiral interface\cite{Wang2018,Mulkers2018}.

\section{Numerical results}\label{sec3}
To confirm our theoretical analysis, we employ full micromagnetic simulations using Mumax3 \cite{Vansteenkiste2014}. We consider a heterogeneous ferromagnetic film with length 5 $\mathrm{\mu{m}}$, width 5 $\mathrm{\mu{m}}$, and thickness 5 nm, in which the DMI strength of the lower half part is $D=3.0$ $\mathrm{mJ/m^{2}}$ while the upper part has a vanishing DMI, as shown in Fig. \ref{fig1}(a). Magnetic parameters of permalloy are used in simulations: $M_{s}=8\times10^{5}$ $\mathrm{A/m}$ and $A=13$ $\mathrm{pJ/m}$. In the dynamic simulations, a Gilbert damping constant of $\alpha=0.001$ is used to ensure a long-distance propagation of the spin-wave beams, and absorbing boundary conditions are adopted to avoid the spin-wave reflection by film edges \cite{Venkat2018}. An external field $\mu_{0}H_{0}=1$ T along $+\hat{y}$ is applied to saturate the in-plane magnetization.

\begin{figure}
  \centering
  \includegraphics[width=0.5\textwidth]{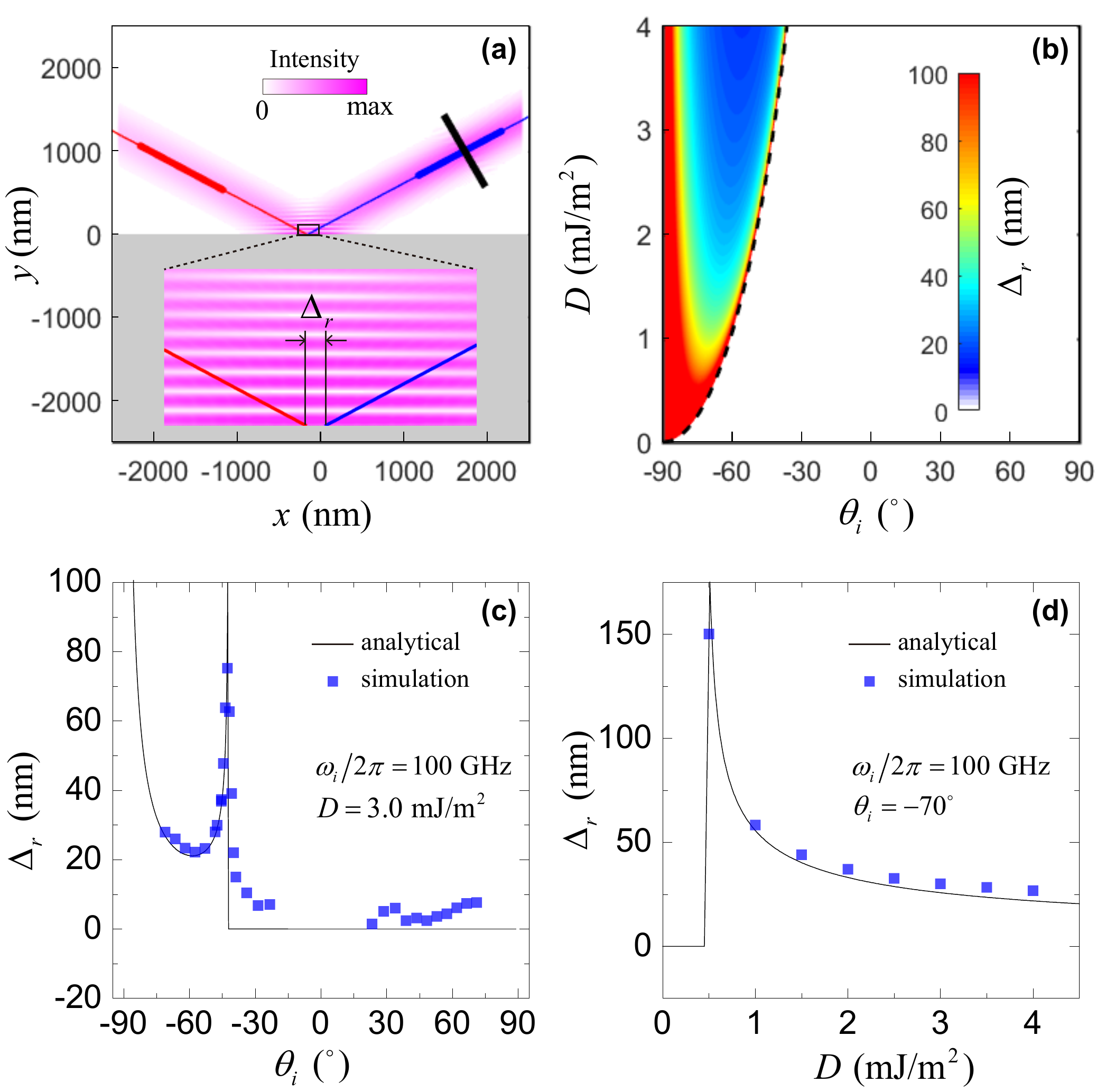}\\
  \caption{(a) Intensity map obtained from the micromagnetic simulations of spin waves reflected from a DMI interface for $\theta_{i}=-60^{\circ}$ and $D=3.0$ $\mathrm{mJ/m^{2}}$. The blue and red lines correspond to the rays of the incident and reflected spin-wave beam, respectively. The black bar denotes the exciting location of spin-wave beams. The inset shows the enlarge image of the rectangular region containing the incident and reflected points. The extracted value of the GH shift of the reflected spin-wave beam is $\Delta_{r}=23.3$ nm. (b) Phase diagram of the GH shift, $\Delta_{r}$, in dependence on $\theta_{i}$ and $D$. Color depicts values of $\Delta_{r}$. The dashed line represents the critical angle $\theta_{ic}$ for different DMI constants at heterochiral interfaces. (c) The GH shift $\Delta_{r}$ as a function of the incident angle $\theta_{i}$ for $D=3.0$ $\mathrm{mJ/m^{2}}$. (d) The dependence of the GH shift $\Delta_{r}$ on the DMI constant for $\theta_{i}=-70^{\circ}$. The spin-wave frequency in (a)-(d) is  $\omega_{i}/2\pi=100$ GHz. In (c) and (d), the blue squares correspond to the simulation data with the demagnetization effect. The solid black line represents the analytical model for the GH shift [Eq. (\ref{eq_GHrTR})].}\label{fig2}
\end{figure}

Next, we excite a Gaussian spin-wave beam by applying a sinusoidal monochromatic microwave field $\mathbf{H}_{\mathrm{ext}}=h_{0}\sin(\omega{t})\hat{z}$ in a narrow rectangular area [black bar shown in Fig. \ref{fig2}(a)], where the field amplitude $h_{0}$ has a Gaussian profile in the transverse direction (parallel with the long side of the exciting area) \cite{Gruszecki2015}. We set the maximum amplitude of the oscillating field as $\mu_{0}h_{0}=10$ mT and $\omega/2\pi=100$ GHz, at which the spin-wave propagation is almost isotropic in the no-DMI region owing to the significant contribution of the exchange interaction [as confirmed in Fig. \ref{fig1}(b)]. It justifies our previous assumption of dropping the dipolar term.

After a sufficiently long-time simulation (3 ns), such that the incident and reflected spin-wave beams are clearly visible, we investigate the GH shift $\Delta_{r}$ of the reflected spin-wave beam. A case study is shown in Fig. \ref{fig2}(a). First, we calculate the spin-wave intensity using equation $I(x,y)=\int_{0}^{t}[\delta{m}_{z}(x,y,t)]^{2}dt$.  Then, by a Gaussian fitting we extract the position of the center of the intensity profile [the blue and red dots in Fig. \ref{fig2}(a) for the incident and reflected spin-wave beams, respectively]. Having the coordinates of the center positions, we obtain the incident and reflected spin-wave beam rays by a linear fitting. Finally, the GH shift of the reflected spin-wave beam $\Delta_{r}$ can be easily calculated [see the inset in \ref{fig2}(a)].

According to the analytical formula, we plot the phase diagram of the GH shift as a function of the incident angle $\theta_{i}$ and DMI constant $D$ as displayed in Fig. \ref{fig2}(b). One can observe that the GH shift is zero at incident angles above the critical angle ($\theta_{i}>\theta_{ic}$) and is finite only if $\theta_{i}<\theta_{ic}$.
Figure \ref{fig2}(c) shows the GH shift $\Delta_{r}$ versus $\theta_{i}$ for $D=3.0$ $\mathrm{mJ/m^{2}}$. The GH shift is pronounced for incident angles close to $\theta_{ic}\approx-42.3^{\circ}$.
With the decreasing of $\theta_{i}$, the GH shift $\Delta_{r}$ first decreases to a minimum value ($\sim21$ nm) at a certain incident angle ($\theta_{i}\approx-58^{\circ}$) and then it starts to increase.
In Fig. \ref{fig2}(d), we show the dependence of the GH shift on the DMI constant for $\theta_{i}=-70^{\circ}$. For $\theta_{i}<\theta_{ic}$ in the total reflection region, the GH shift decreases with $D$.
The simulation data are represented by the blue squares in Figs. \ref{fig2}(c) and \ref{fig2}(d).
These results are well consistent with those obtained from the analytical model Eq. (\ref{eq_GHrTR}).

However, there are still some discrepancies which are possibly due to approximations in the analytical model, as discussed below. Firstly, the demagnetization field is considered in simulations, which is neglected in the analytical model. With the demagnetization field, the magnitude of the wave vector $k_{r}^{0}$ is reduced [see Fig. \ref{fig1}(b)], leading to the increase of $\Delta_{r}$ in Eq. (\ref{eq_GHrTR}) [as shown in Figs. \ref{fig2}(c) and \ref{fig2}(d)]. Secondly, the magnetization saturation is assumed in the analytical model. But there is spin canting at the DMI interface \cite{Mulkers2017}, which would change the spin-wave dispersion \cite{Wang2018}. And it would cause bending of spin-wave beams at the interface, which is similar to the spin-wave bending due to the gradual change of refractive index caused by the demagnetization field at the film edge \cite{Gruszecki2014}. Lastly, for $\theta_{i}>\theta_{ic}$, most of the incident spin-wave beams pass through the interface. The intensity of the reflected beam is very weak leading to sizable fitting errors.

\begin{figure}
  \centering
  \includegraphics[width=0.5\textwidth]{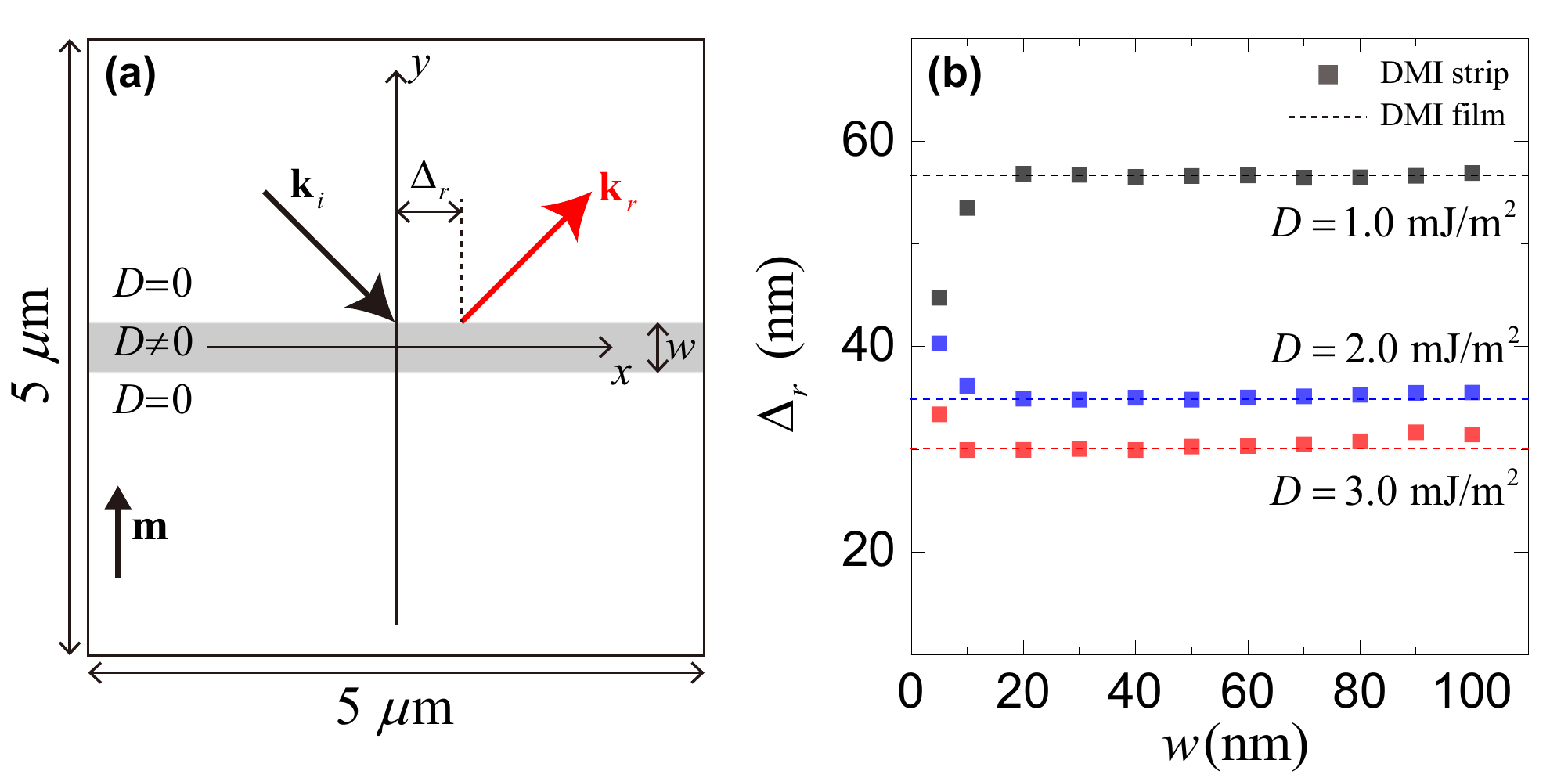}\\
  \caption{(a) Schematic plot of the GH shift $\Delta_{r}$ at the DMI interface in a heterogeneous film with a DMI strip (gray region) in the center. The width of the strip is $w$. (b) The simulation results of the GH shift $\Delta_{r}$ in dependence on the strip width $w$ for three DMI constants. The incident angle and frequency are $\theta_{i}=-70^{\circ}$ and $\omega_{i}/2\pi=100$ GHz, respectively. The dashed lines and squares correspond to the GH shift at the interface of DMI film and DMI strip, respectively.}\label{fig3}
\end{figure}

Finally, we replace the DMI film by the DMI strip and investigate the influence of the strip width on the GH shift. To this end, we construct a heterogeneous ferromagnetic film, in which a DMI strip with the finite width is located in the center while the remaining parts have no DMI, as shown in Fig. \ref{fig3}(a). Then we numerically examine how the GH shift change with the strip width. Figure \ref{fig3}(b) shows the simulation results for $\theta_{i}=-70^{\circ}$, $\omega_{i}/2\pi=100$ GHz, and three different DMI constants.
One can see that the GH shift is independent of the strip width and remains the same as in the DMI film for $w>10$ nm. For $w<10$ nm, there is a large variation of the GH shift. We contribute this change to the spin canting, the range of which is comparable with the strip width. In this case, the dispersion relation Eq. (\ref{eq_dispersion}) and boundary conditions Eq. (\ref{eq_bc}) become invalid. The incident spin-wave beams partially pass though the interface rather than being totally reflected (not shown).
A new model is required to elucidate the GH shift at a ultra-narrow DMI interface ($w<10$ nm), which goes beyond the scope of this work.
\section{Discussion}\label{sec4}
We point out that the results presented here enable us to probe the DMI of magnetic strip with the width of sub-$50$ nm by measuring the GH shift between the incident and reflected spin-wave beams. This method requires the spin-wave imaging with the spatial resolution in the range of about 500 nm (the width of the spin-wave beam), which can be achieved by X-ray magnetic circular dichroism (XMCD) \cite{Bonetti2015} or the near-field BLS \cite{Jersch2010}. Recently, a GH-like phase shift for magnetostatic spin waves has been observed in experiments \cite{Stigloher2018}. We thus envision a direct observation of the GH shift of exchange spin waves in the future.

In previous work, we proposed a nonlocal scheme to measure the DMI in a narrow magnetic strip by three-magnon processes \cite{Wang2018}. Such a method is only feasible for the magnetic strip with the width in the range of 50-100 nm. Below 50 nm, the spin canting \cite{Mulkers2017} at the DMI interface elevates the band gap of bound-state spin waves in the DMI strip leading to a very narrow frequency range to excite the confined modes. This significantly hiders the measurement of DMI. Beyond 100 nm, the three-magnon intensity attenuates rapidly with the strip width. It is thus difficult to detect the spectra of the transmitted spin waves. We therefore believe that GH shift provides a very promising way to quantify the DMI in ultra-narrow magnetic strips.

\section{Conclusion}\label{sec5}
In summary, we investigated the GH shift at the interface between two ferromagnets with different DMIs. The analytical formulas for the reflectance and the GH shift were derived. We have shown that the GH shift of the reflected spin-wave beam only takes place in the case of total reflection and disappears for the other cases. Using the micromagnetic simulations, we demonstrated that the analytical model well describes the essential physics. Further, we studied the GH shift at the interface of a DMI strip and found that the GH shift is independent on the strip width for $w>10$ nm. Our findings are helpful to understand the GH shift in chiral magnets and to measure the DMI parameter in ultra-narrow magnetic strips.

\section{Acknowledgment}
We thank C. Wang, Z.-X. Li, and H. Yang for helpful discussions.
This work is supported by the National Natural Science Foundation of China (Grants No. 11604041 and 11704060), the National Key Research Development Program under Contract No. 2016YFA0300801, and the National Thousand-Young-Talent Program of China. Z.W. acknowledges the financial support from the China Postdoctoral Science Foundation under Grant No. 2019M653063.


\end{document}